\RequirePackage{fix-cm}
\documentclass[smallextended]{svjour3}       % onecolumn (second format)
\smartqed  % flush right qed marks, e.g. at end of proof
\usepackage{graphicx}
\usepackage{bm}  
\usepackage{amsmath}  
\begin{document}

\title{Dynamics of a massive intruder in a homogeneously driven granular
fluid%\thanks{Grants or other notes
}


\author{A. Puglisi \and A. Sarracino \and G. Gradenigo \and D. Villamaina}


\institute{All authors \at
CNR-ISC c/o Dip. di Fisica, \\
Universita` degli Studi di Roma "La Sapienza", \\
Piazzale Aldo Moro 2, I-00185 Roma, Italy. \\
Tel.:+39-0649913508 Fax:+39-064463158 \\
\email{andrea.puglisi@roma1.infn.it}             \\
\email{ale.sarracino@gmail.com} \\
\email{ggradenigo@gmail.com}     \\        
\email{villamaina@gmail.com}
}

\date{Received: date / Accepted: date}

\maketitle

\begin{abstract}
A massive intruder in a homogeneously driven granular fluid, in dilute configurations, performs
a memory-less Brownian motion with drag and temperature simply related
to the average density and temperature of the fluid. At volume fraction
$\sim 10-50\%$ the intruder's velocity correlates with the
local fluid velocity field: such situation is approximately described by a system of coupled
linear Langevin equations equivalent to a generalized Brownian motion
with memory. Here one may verify the breakdown of the
Fluctuation-Dissipation relation and the presence of a net entropy
flux - from the fluid to the intruder - whose fluctuations satisfy the
Fluctuation Relation.

\end{abstract}

%\section{Introduction}
%\label{intro}

Granular fluids represent a valid benchmark for modern theories of
non-equilibrium statistical mechanics~\cite{JNB96b}. Due to
dissipative interactions among the microscopic constituents, energy is
not conserved and an external source is necessary to maintain a
stationary state. The consequence is a breakdown of time reversal
invariance and the failure of properties such as the Equilibrium
Fluctuation-Dissipation relation (EFDR)~\cite{GN00}. In recent years,
a systematic theory for the dilute limit has been developed, in good
agreement with numerical simulations~\cite{BP04,BMG09}, while a
general understanding of dense granular fluids is still lacking. A
common approach is the so-called Enskog correction~\cite{BP04,DS06},
which reduces the breakdown of Molecular Chaos to a renormalization of
the collision frequency. In cooling regimes, the Enskog theory may
describe strong non-equilibrium effects, due to the explicit cooling
time-dependence~\cite{SD01}. Nevertheless it cannot describe dynamical
effects in stationary regimes, such as multiple characteristic times or different decays
of response and autocorrelation~\cite{G04,PBV07}.

Here we review a recent model~\cite{SVGP10} for the dynamics of a
massive tracer moving in a gas of smaller granular particles, both
coupled to an external bath. Taking as reference point the dilute
limit, where the system has a closed analytical
description~\cite{SVCP10}, a Langevin equation linearly coupled to a fluctuating local velocity field is
proposed as first approximation capable of describing the dense
case. Its main features are: i) the decay of correlation and response
functions is not simply exponential and shows
backscattering~\cite{OK07,FAZ09} and ii) the EFDR~\cite{KTH91,BPRV08}
of the first and second kind do not hold.  In such a model, detailed
balance is not necessarily satisfied, and a fluctuating entropy
production~\cite{seifert05} can be measured, which fairly verifies the
Fluctuation Relation~\cite{Kurchan,LS99,BPRV08}. 
%This is a remarkable
%result, if considered the interest of the community~\cite{BGGZ06b} and
%compared with unsuccessful past attempts~\cite{FM04,PVBTW05}.

%\section{Model}
%\label{Model}

The model reviewed here is the following: an ``intruder'' disc of mass $m_0=M$ and radius $R$,
moving in a gas of $N$ granular discs with mass $m_i=m$ ($i>0$) and
radius $r$, in a two dimensional box of area $A=L^2$. We denote by
$n=N/A$ the number density of the gas and by $\phi$ the occupied
volume fraction, i.e. $\phi=\pi(Nr^2+R^2)/A$ and we denote by $\bm{V}$
(or $\bm{v}_0$) and $\bm{v}$ (or $\bm{v}_i$ with $i>0$) the velocity
vector of the tracer and of the gas particles, respectively.
Interactions among the particles are hard-core binary instantaneous
inelastic collisions, such that particle $i$, after a collision with
particle $j$, comes out with a velocity
$\bm{v}_i'=\bm{v}_i-(1+\alpha)\frac{m_j}{m_i+m_j}[(\bm{v}_i-\bm{v}_j)\cdot\hat{\bm{n}}]\hat{\bm{n}}$
where $\hat{\bm{n}}$ is the unit vector joining the particles' centers
of mass and $\alpha \in [0,1]$ is the restitution coefficient
($\alpha=1$ is the elastic case).
The mean free path of the intruder is proportional to $l_0=1/(n(r+R))$ and we denote
by $\tau_c$ its mean collision time. Two kinetic temperatures can be
introduced for the two species: the gas temperature
$T_g=m\langle \bm{v}^2\rangle/2$ and the tracer temperature
$T_{tr}=M\langle \bm{V}^2\rangle/2$.

The equation of motion of the $i$-th particle reads:
$m_i\dot{v}_i(t)=-\gamma_b v_i(t) + f_i(t) + \xi_b(t)$.
%\label{langgas}
Here $f_i(t)$ is the force taking into account the collisions of particle $i$ with other
particles, and $\xi_b(t)$ is a white noise (different for all particles), with
$\langle\xi_b(t)\rangle=0$ and $\langle\xi_{b}(t)\xi_{b}(t')\rangle
=2T_b\gamma_b\delta(t-t')$.  The effect of the external energy source
balances the energy lost in the collisions and a stationary state is
attained with $m_i\langle v_i^2 \rangle \leq T_b$~\cite{WM96,NETP99,PLMPV98,PEU02,GSVP11b} .

At low packing fractions, $\phi < 0.1$, and in the large mass limit,
$m/M\ll 1$, using the Enskog approximation it has been
shown~\cite{SVCP10} that the dynamics of the intruder is described by
a linear Langevin equation: 
\begin{equation} \label{lange}
M\dot{V}=-\Gamma_E V+ {\cal E}_E,
\end{equation}
with ${\cal E}_E$ a white noise with $\langle {\cal E}_E \rangle =0$,
$\langle{\cal E}_E(t){\cal E}_E(t')\rangle=2\delta(t-t')\Gamma_E
T_{tr}^E$ and
$T_{tr}^E=(\gamma_bT_b+\gamma_g^E\frac{1+\alpha}{2}T_g)/\Gamma_E$ is
the tracer's temperature.  In this limit the velocity autocorrelation
function shows a simple exponential decay, with characteristic time
$M/\Gamma_E$, where $\Gamma_E=\gamma_b+\gamma_g^E$ and
$\gamma_g^E=\frac{g_2(r+R)}{l_0}\sqrt{2\pi mT_g}(1+\alpha)$ where
$g_2(r+R)$ is the pair correlation function for a gas particle and the
intruder at contact. Time-reversal and the EFDR, weakly modified for
uniform dilute granular gases~\cite{PBL02,G04,PVTW06}, become
perfectly satisfied for a massive intruder.

As the packing fraction is increased, the Enskog approximation fails
in predicting dynamical properties.  In particular, velocity
autocorrelation $C(t)=\langle V(t)V(0)\rangle/\langle V^2\rangle$ and
linear response function $R(t)=\overline{\delta V(t)}/\delta V(0)$
show an exponential decay modulated by oscillating
functions~\cite{FAZ09,SVGP10}.  Moreover violations of the EFDR
$C(t)=R(t)$ are observed for $\alpha<1$~\cite{PBV07,VPV08}.  The
Enskog approximation is unable to explain the observed functional
forms, because it only modifies by a constant factor the collision
frequency~\cite{BP04,SVCP10}: a model with more than one
characteristic time is needed.  A first approximation is given by an
auxiliary field coupled to the intruder's velocity:
%(see~\cite{VBPV09,SVGP10} for details on how the mapping is achieved):
\begin{eqnarray} \label{local_field}
M\dot{V}=-\Gamma_E(V-U)+\sqrt{2\Gamma_E T_g}{\cal E}_V\\ \nonumber
M'\dot{U}=-\Gamma' U - \Gamma_E V+\sqrt{2 \Gamma' T_b} {\cal E}_U,
\end{eqnarray}
where ${\cal E}_V$ and ${\cal E}_U$ are white noises of unitary
variance. Two new parameters appear: the mass of the local field $M'$
and its drag coefficient $\Gamma'$. The dilute limit here is obtained
for $\Gamma' \sim M' \to \infty$. In such a limit indeed $U \to 0$ and
the equation for $V$ comes back in the form discussed
above~\cite{SVCP10}.  In such a form~(\ref{local_field}), the dynamics
of the tracer is remarkably simple: indeed $V$ follows a Langevin
equation in a {\em Lagrangian frame} with respect to a field $U$,
which is the \emph{local average velocity field} of the gas particles
colliding with the tracer. A first justification of this model comes
from realizing~\cite{SVGP10} that it is equivalent to a Generalized
Langevin Equation with exponential memory, which is consistent with a
typical approximation done for Brownian Motion when, at high
densities, the coupling of the intruder with fluid hydrodynamic modes,
decaying exponentially in time~(see \cite{Z01}, Cap. 8.6 and 9.1),
must be taken into account. Such a coupling, which in principle
involves a continuum of modes, is reduced here to a single dominant
mode: this is sufficient to introduce a new non-trivial timescale. The full
coupling would reproduce finer features which become relevant at
larger densities or larger times, such as long-time power-law
tails. The fact that the ``temperature'' of the local velocity field
$U$ is equal to the bath temperature $T_b$ comes as a consequence of
the conservation of momentum in collisions, implying that the average
velocity of a group of particles is not changed by collisions among
themselves and is only affected by the external bath and a (small)
number of collisions with outside particles. This scenario is fully
consistent with recent study of hydrodynamic fluctuations for the
velocity field of the same fluid model~\cite{GSVP11,GSVP11b}.

\begin{figure*}[!htb]
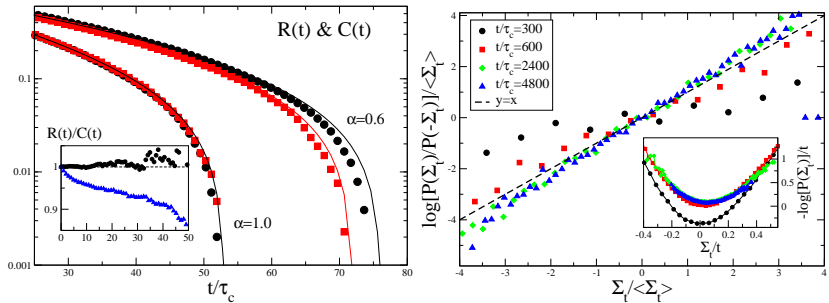

\includegraphics[width=.45\columnwidth,clip=true]{figura2.eps}
\includegraphics[width=.45\columnwidth,clip=true]{figura3.eps}
\caption{(Color online). Left: correlation function $C(t)$ (black circles) and response function $R(t)$ (red
squares) for $\alpha=1$ and $\alpha=0.6$, at $\phi=0.33$. 
Continuous lines show curves obtained with Eqs.~(\ref{Cfit}). 
Inset: the ratio $R(t)/C(t)$ is reported in the same cases. Right: Check of the fluctuation
relation~(\ref{GCrel}) in the system with $\alpha=0.6$ and
$\phi=0.33$. 
Inset: collapse of the rescaled probability
distributions of $\Sigma_t$ at large times onto the large deviation function.}
\label{fig_resp}
\end{figure*}

A stronger justification comes, however, from its effectivness in reproducing
the numerical results, as detailed in~\cite{SVGP10}. From
the simulations it is seen that the relaxation time of the local field
$\tau_U=M'/\Gamma'$, rescaled by the mean collision time, increases with the packing fraction and with the
inelasticity, as expected. At high densities it appears that
$\Gamma' \sim 1/\phi$, and $T_{tr} \sim T_g \sim
T_g^E$, likely due to stronger correlations among
particles. At large $\phi$ we observe $T_{tr}>T_{tr}^E$,
consistent with a {\em smaller} dissipation for correlated collisions. Model~(\ref{local_field}) predicts
$C=f_C(t)$ and $R=f_R(t)$ with
\begin{equation}
f_{C(R)}=e^{-gt}[\cos(\omega t)+a_{C(R)}\sin(\omega t)]. \label{Cfit} 
\end{equation} 
The variables $g$, $\omega$, $a_C$ and $a_R$ are known algebraic
functions of $\Gamma_E$, $T_g$, $\Gamma'$, $M'$ and $T_b$. In
particular, the ratio $a_C/a_R=[T_g-\Omega (T_b-T_g)] /[T_g+\Omega
  (T_b-T_g)]$, with
$\Omega=\Gamma_E/((\Gamma'+\Gamma_E)(\Gamma_EM'/M-\Gamma'))$.  Hence,
in the elastic ($T_g \to T_b$) as well as in the dilute limit
($\Gamma' \to \infty$), one gets $a_C=a_R$ and recovers the EFDR
$C(t)=R(t)$. Such predictions are all verified in numerical simulations~\cite{SVGP10}. In particular 
Fig.~\ref{fig_resp} depicts correlation
and response functions in a dense case (elastic and inelastic):
symbols correspond to numerical data and continuous lines the analytical
curves.  In the inelastic case, deviations from EFDR $R(t)=C(t)$
are observed. In the inset of Fig.~\ref{fig_resp} the ratio
$R(t)/C(t)$ is also reported.  It is important to notice that the main responsibility for the breakdown of the EFDR is the coupling between $V$ and $U$, indeed Eq.~\eqref{Cfit} can be expressed in a different way: $R(t)=a C(t) + b \langle V(t)U(0)\rangle$
with $a=[1-(T_g-T_b)\Omega_a/\Gamma']$ and $b=(T_g-T_b)\Omega_b$,
where $\Omega_a$ and $\Omega_b$ are known functions of the parameters. At 
equilibrium or in the dilute limit the EFDR is recovered.

An important independent assessment of model~\eqref{local_field} comes from the study of the fluctuating
entropy production~\cite{seifert05} which quantifies the deviation
from detailed balance in a trajectory. Given the trajectory in the
time interval $[0,t]$, $\{V(s)\}_0^t$, and its time-reversed $\{{\cal
  I}V(s)\}_0^t\equiv\{-V(t-s)\}_0^t$, the entropy production
for our model takes the form~\cite{PV09}
\begin{equation}
\Sigma_t=\log\frac{P(\{V(s)\}_0^t)}{P(\{{\cal I}V(s)\}_0^t)}
\approx \Gamma_E\left(\frac{1}{T_g}-\frac{1}{T_b}\right)\int_0^t ds~V(s)U(s).
\label{entropy_prod}
\end{equation}
This functional vanishes exactly in the
elastic case, $\alpha=1$, where equipartition holds, $T_g=T_b$, and is
zero on average in the dilute limit, where $\langle VU\rangle=0$.
Formula~\eqref{entropy_prod} reveals that the leading source of
entropy production is the energy transferred by the ``force''
$\Gamma_E U$ on the tracer, weighed by the difference between the
inverse temperatures of the two ``thermostats''. Therefore, to measure
entropy production, we need to measure the fluctuations of $U$: a possible choice is a local average of particles'
velocities in a circle of radius $l+R$ centered on the
tracer. Details on how to choose in a reliable way the correct $l$ are
given in~\cite{SVGP10}. Following such procedure, in the case
$\phi=0.33$ and $\alpha=0.6$, we estimate for the
correlation length $l \sim 9r \sim 6l_0$. Then, measuring 
 the entropy production from Eq.~(\ref{entropy_prod}) 
along many trajectories of length $t$, we computed the probability $P(\Sigma_t=x)$ and compared it to
$P(\Sigma_t=-x)$, in order to verify the Fluctuation Relation~\cite{Kurchan,LS99,BPRV08}
\begin{equation}
\log\frac{P(\Sigma_t=x)}{P(\Sigma_t=-x)}=x.
\label{GCrel}
\end{equation}
In the right frame of Fig.~\ref{fig_resp} the
results of this comparison are reported. The main frame confirms that at large times the Fluctuation
Relation~(\ref{GCrel}) is well verified within the statistical errors.
The inset shows the collapse of $\log P(\Sigma_t)/t$ onto the large
deviation rate function for large times.  Notice that - in
formula~\eqref{entropy_prod} - a wrong evaluation of the weighing
factor $(1/T_g-1/T_b)$ or of the ``energy injection rate'' $\Gamma_E
U(t) V(t)$ in Eq.~\eqref{entropy_prod} could produce a completely
different slope in Fig.~\ref{fig_resp} (right frame).

To conclude this paper, we stress that velocity correlations
$\langle V(t) U(t') \rangle$ between the intruder and the surrounding
velocity field are responsible for both the violations of the EFDR and
the appearance of a non-zero entropy production, provided that the two
fields are {\em at different temperatures}.  We also mention that
larger violations of EFDR can be observed using an intruder with a
mass equal or similar to that of other particles~\cite{PBV07}, with
the important difference that in such a case a simple
``Langevin-like'' model for the intruder's dynamics is not available.

%\begin{figure}
%  \includegraphics{example.eps}
%\end{figure}
%\begin{figure*}
%  \includegraphics[width=0.75\textwidth]{example.eps}
%\end{figure*}
%\begin{table}
%\begin{tabular}{lll}
%\hline\noalign{\smallskip}
%\end{tabular}
%\end{table}

\begin{acknowledgements}
This work is dedicated to the memory of Isaac Goldhirsch, from whom we
learned plenty of science, but also a smiling attitude toward serious things. 
The work is  supported by the ``Granular-Chaos'' project, MIUR grant number RBID08Z9JE.
\end{acknowledgements}

\bibliographystyle{spmpsci}      % mathematics and physical sciences
\bibliography{fluct.bib}   % name your BibTeX data base

%\begin{thebibliography}{}
% etc
%\end{thebibliography}

\end{document}